\documentclass[usenatbib,useAMS,preprint,psfig2]{mn2e}
\usepackage{times}
\usepackage{amsmath}
\usepackage{amssymb,amsfonts}
\usepackage{graphicx}
\usepackage{epstopdf}
\usepackage{verbatim}

\def\gapprox{\lower.4ex\hbox{$\;\buildrel >\over{\scriptstyle\sim}\;$}}
\def\lapprox{\lower.4ex\hbox{$\;\buildrel <\over{\scriptstyle\sim}\;$}} \def\be{\begin{equation}}
\def\be{\begin{equation}}
\def\ee{\end{equation}}
\def\bea{\begin{eqnarray}}
\def\eea{\end{eqnarray}}
\catcode`\@=11
\font\tenmib=cmmib10
\font\tensyb=cmbsy10
\font\tenbi=cmmib10 
\newfam\bifam  
\textfont\bifam=\tenbi  %for bold italics; textstyle only

\def\unboldmath{\everymath{}\everydisplay{}
          \textfont\@ne\teni 
          \textfont\tw@\tensy
          }
\def\boldmath{$\!\!$\relax\everymath{\mit}\everydisplay{\mit}
        \textfont\@ne\tenmib
        \textfont\tw@\tensyb 
        \relax}%
\catcode`\@=12
  
\begin{document}
\title[Anisotropic Scintillations]{Extremely Anisotropic Scintillations}
\author[Walker,  de~Bruyn \&\ Bignall]{M.A.~Walker$^1$, A.G. de Bruyn$^{2,3}$, H.E. Bignall$^{4,5}$\\
1. Manly Astrophysics Workshop, Unit 3, 22 Cliff Street, Manly, NSW 2095, Australia\\
2. Netherlands Foundation for Research in Astronomy, P.O. Box 2, 7990 AA Dwingeloo, The Netherlands\\
3. Kapteyn Astronomical Institute, University of Groningen, P.O. Box 800, 9700 AV Groningen,
The Netherlands\\
4. Joint Institute for VLBI in Europe, Postbus 2, 7990AA Dwingeloo, The Netherlands\\
5. Curtin Institute of Radio Astronomy, Curtin University of Technology,
GPO Box U1987, Perth, WA 6845, Australia}

\date{\today}

\maketitle

\begin{abstract}
A small number of quasars exhibit interstellar scintillation on time-scales less than an hour; their scintillation patterns are all known to be anisotropic. Here we consider a totally anisotropic model in which the scintillation pattern is effectively one-dimensional.  For the persistent rapid scintillators J1819$+$3845 and PKS1257$-$326 we show that this model offers a good description of the two-station time-delay measurements and the annual cycle in the scintillation time-scale. Generalising the model to finite anisotropy yields a better match to the data but the improvement is not significant and the two additional parameters which are required to describe this model are not justified by the existing data. The extreme anisotropy we infer for the scintillation patterns must be attributed to the scattering medium rather than a highly elongated source. For J1819$+$3845 the totally anisotropic model predicts that the particular radio flux variations  seen between mid July and late August should repeat between late August and mid November, and then again between mid November and late December as the Earth twice changes its direction of motion across the scintillation pattern.  If this effect can be observed then the minor-axis velocity component of the screen and the orientation of that axis can both be precisely determined. In reality the axis ratio is finite, albeit large, and spatial decorrelation of the flux pattern along the major axis may be observable via differences in the pairwise fluxes within this overlap region; in this case we can also constrain both the major-axis velocity component of the screen and the magnitude of the anisotropy.
\end{abstract}


\section{Introduction}
There are three quasars which are known to exhibit large-amplitude radio flux variations on time-scales of less than an hour: PKS0405$-$385 (Kedziora-Chudczer et al 1997), whose variations are intermittent, and the persistent variables PKS1257$-$326 (Bignall et al 2003) and J1819$+$3845 (Dennett-Thorpe \&\ de~Bruyn 2000).  The short time-scale and the strong frequency dependence of the observed variations indicate that interstellar scintillation is the cause (Kedziora-Chudczer et al 1997). This conclusion is compellingly reinforced by the observed annual cycle in variability time-scale which is effected by the rotation of the Earth's orbital velocity vector (Dennett-Thorpe \&\ de~Bruyn 2003; Bignall et al 2003). Similarly the two-station time-delay measurements (Dennett-Thorpe \&\ de~Bruyn 2002; Bignall et al 2006) explicitly demonstrate the spatial modulation of flux which is inherent in scintillation. The annual cycle in the characteristic time-scale of the scintillations provides strong constraints on the velocity of the scattering material; in turn this constrains the distance to the scattering screen if one identifies the scintillation time-scale with the time taken to cross a Fresnel zone. Better constraints on the screen distance are obtained by constructing models and attempting to match them to all aspects of the data for a given source; the distances deduced in this way are surprisingly small -- of order 10pc from Earth (Dennett-Thorpe \&\ de~Bruyn 2003; Bignall et al 2003) -- so very high brightness temperatures are not required for the sources. 

The same annual cycle in the scintillation time-scale also provides evidence for strong anisotropy in the scintillation pattern (Dennett-Thorpe \&\ de~Bruyn 2003; Bignall et al 2003). The scintillation time-scale is affected by both anisotropy in the scattering material and in the source structure. Quasars often exhibit elongated structure (``jets'') on milliarcsecond and larger scales (e.g. Walker, Benson \&\ Unwin 1987) so it is likely that they do have elongated structure on the sub-milliarcsecond angular scales relevant to interstellar scintillation. However, the spectral purity of the observed light-curves -- i.e. their quasi-sinusoidal nature -- argues that the scattering screens must also have anisotropic structure (Rickett, Kedziora-Chudczer \&\ Jauncey 2002; see also Bignall et al 2003), and that the major-to-minor axis ratio  is large ($> 4$). Unfortunately the spectral purity of the light-curves does not offer a sensitive test of the level of anisotropy once the axis ratio becomes large. 

Because they are so nearby, yet they produce large amplitude variations, we know that the screens reponsible for intra-hour variability in quasars must be regions of very strong scattering -- i.e. they must scatter radio-waves through large angles -- and if the local interstellar medium is representative then similar, more distant screens could make a substantial contribution to the total scattering seen on other lines-of-sight. For compact radio quasars this suggests that smaller amplitude variability on longer time-scales should be relatively common and this expectation is qualitatively consistent with the results of flux monitoring of large numbers of compact extragalactic radio sources (Lovell et al 2003).

A further indication that similar screens are widely distributed in the Galaxy is the observation of  parabolic arcs in the ``secondary spectra'' (power spectra of the dynamic spectra) of radio pulsars (Stinebring et al 2001). Modelling of this phenomenon indicates that localised, strongly-scattering screens are responsible, that the scattering is anisotropic and that the screens are  hundreds of parsec distant from the Sun (Stinebring et al 2001; Cordes et al 2006; Walker et al 2004; Putney \&\ Stinebring 2006). It therefore seems that the three known intra-hour variable quasars can offer insights into the structure of the broader interstellar medium and it is important that we understand as much as we can about the properties of these nearby scattering screens.

As we have just noted, it is already clear that the major-to-minor axis ratio of the scintillation patterns is large for the three intra-hour variable quasars. Here we seek to determine whether existing data admit the possibility of the patterns being so anisotropic that the behaviour is effectively one-dimensional and, if so, whether that is the preferred model. In \S2 we use published data on  the annual cycle in scintillation time-scale and the two-station time-delay experiments for both J1819$+$3845 and PKS1257$-$326 to test our models. We find that infinite anisotropy in the scintillation patterns is  consistent with and slightly preferred by existing data, and in \S3 we interpret this result as one-dimensional scattering rather than the influence of anisotropy in the source structure. In \S3 we further note that extreme flux-pattern anisotropy leads us to expect the J1819$+$3845 light-curves seen at one time of year to repeat at two other times, as the observer moves back-and-forth across the same region of the scintillation pattern. If the major-to-minor axis ratio of the pattern is $R\la10^5$ then spatial  decorrelation along the major-axis direction may be measurable and flux monitoring of J1819$+$3845 can then be used to determine the length-scales and velocity components of the screen along both major-  and minor-axis directions. 

\section{Model fitting}
We attempted to fit the data for each of the persistent scintillators with both one- and two-dimensional scintillation  models. Our two-dimensional model is the one given in Bignall et al (2006). Five free parameters are needed to characterise the model, they are: the orientation angle, $\beta$, of the major axis of the flux-pattern anisotropy; the steady velocity, $\vec{v}$, with components $v_\parallel, v_\perp$ parallel and perpendicular, respectively, to this major axis direction; the characteristic length-scale, $a_\perp$, of the flux pattern measured along its minor axis; and the ratio, $R$, of major-to-minor axis length scales ($R=a_\parallel/a_\perp$). Henceforth we use the $\parallel, \perp$ notation to indicate any vector  components resolved onto the major and minor axes of the scintillation pattern; these axes lie in the plane perpendicular to the direction to the source.

The orbital velocity of the Earth, $\vec{u}_\oplus$, contributes to the total effective velocity of the observer across the scintillation pattern $\vec{V}:=\vec{u}_\oplus + \vec{v}$, and in terms of these quantities the annual cycle in scintillation time-scale, $t_s$, expected in any model is given by
\be
t_s = {{ R a_\perp} \over { \sqrt{ V_\parallel^2 + R^2 V_\perp^2} } },
\ee
and the time-delay measured between two stations separated by a baseline $\vec{b}$ is
\be
\tau = {{b_\parallel V_\parallel + R^2 b_\perp V_\perp} \over {V_\parallel^2 + R^2 V_\perp^2} }.
\ee
Our one-dimensional scattering model corresponds to the limit of infinite axis ratio ($R\rightarrow\infty$) and is completely specified by three of these five parameters: $\beta, v_\perp, a_\perp$. The time-scale and two-station delay in this model are just $t_s=a_\perp/|V_\perp|$ and $\tau=b_\perp/V_\perp$. 

\begin{figure}
\vskip -1.5cm\hskip -1.5cm\includegraphics[width=11cm]{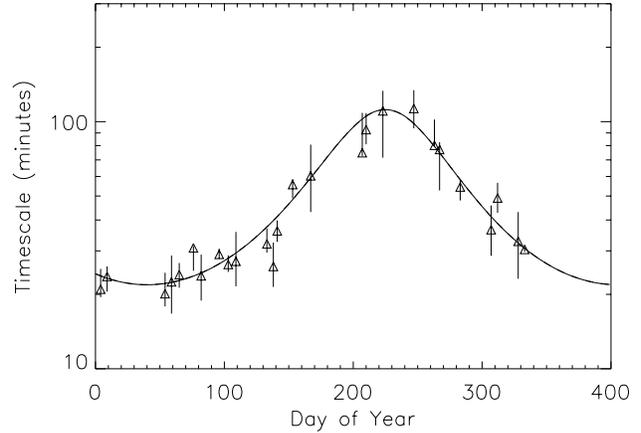}
\vskip -6.5cm
\caption{The annual cycle in scintillation time-scale for PKS1257$-$326 for best-fit one- and two-dimensional models along with the data. The two models are so similar in their predictions that they are indistinguishable here.}
\end{figure}

\subsection{PKS1257$-$326}
The time-scale and delay data for this source are presented in Bignall et al (2006). We have confined our attention to the 4.8~GHz data as these are expected to be less affected by atmospheric phase and opacity fluctuations than the 8.5~GHz data. As in Bignall et al (2006) we fit simultaneously to the time-scale and delay, in order to determine which model best fits the data overall, but our approach differs in two respects from theirs. First, for the two-station experiments we tested our models against the light-curves as measured at the two separate telescopes - the Australia Telescope Compact Array and the Very Large Array - rather than using the mean delay measurements reported in Bignall et al (2006). We adopted this approach because the delay varies during the course of the experiment, as the baseline between the telescopes rotates; this information is in the original light-curves but not in the reported mean delay. However, this refinement is not expected to yield a major improvement in fitting because Bignall et al (2006) looked for the effect by splitting up their data into smaller intervals and found no significant variation of delay during the course of the experiment. The second point of difference relates to how we weight the delay measurements, relative to the scintillation time-scale measurements, in their contribution to the quality-of-fit estimator, $\chi^2$. Bignall et al (2006) weighted the set of time-scale data equally with the set of delay measurements, whereas here we weight each of the 26 measurements of time-scale equally with each of the 5 two-station delay measurements  --- reflecting the fact that similar amounts of data contributed to each measurement. 

\begin{table}
  \begin{minipage}{120mm}
 \caption{Best fit scintillation model parameters (two- and one-dimensional).}
  \begin{tabular}{@{}llllllll@{}}
  \hline
   Source &$\beta$&$v_\perp$&$\!a_\perp$&$v_\parallel$&$R$&$\chi^2$\\
        &${\rm N\!\rightarrow\!E}$&${\rm km\,s^{-1}}$&$\!{\rm Mm}$&${\rm km\,s^{-1}}$ \\
 \hline
1257$-$326&$124.9^\circ$&$19.3$&$\!42.8$&$184$&$385$&$27.6$\\
1257$-$326&$124.9^\circ$&$19.3$&$\!42.8$&$-$&$-$&$27.6$\\
& \\
1819$+$3845&$-97.4^\circ$&$19.6$&$\!29.8$&$400$\footnote{Maximum physically plausible value}&$176$&$50.3$\\
1819$+$3845&$-97.3^\circ$&$19.7$&$\!29.5$&$-$&$-$&$52.6$\\
\hline
\end{tabular}
\end{minipage}
\end{table}

Our best fits to these data, using both one- and two-dimensional models, are described by the parameters given in table 1. Our best-fit one-dimensional model exhibits similar properties to that of the two-dimensional model in respect of the three parameters which are common between the two models. The annual cycle in scintillation time-scale as predicted by each of the best-fit models is shown in figure 1, along with the data; the two models are indistinguishable to the eye and both match the data well. It is inevitable that the $\chi^2$ value for the best-fit two-dimensional model is smaller than that of the best-fit one-dimensional model, just because the two dimensional model has additional parameters which can be adjusted to achieve a better match to the data, but in the present case the two $\chi^2$ values are so close that they are identical to the accuracy given in table 1. Clearly the best two-dimensional model is not significantly better than the best one-dimensional model. This point is emphasised when we normalise our model $\chi^2$ values by the number of degrees of freedom. There are 26 time-scale measurements and 5 two-station delay measurements, so 31 constraints in total. The two-dimensional model contains 5 free parameters whereas the one-dimensional model has only 3, implying reduced-$\chi^2$ values of $\chi_r^2=0.99$ and $\chi_r^2=1.06$ for one- and two-dimensional models, respectively:  the one-dimensional fit is slightly better. This tells us that there is no justification in these data for generalising the one-dimensional model to two-dimensions.

\subsection{J1819$+$3845}
For J1819$+$3845 we use the two station delay observations from January 2001 presented in Dennett-Thorpe \&\ de~Bruyn (2002), and the data on annual time-scale variations (time-scale ``$t_1$'') given in table 3 of Dennett-Thorpe \&\ de~Bruyn (2003). In both of their two-station experiments Dennett-Thorpe \&\ de~Bruyn (2002) clearly measured a change in delay as the VLA-Westerbork baseline rotated during the course of their observations and the sign of the delay changed. We used the measurements as plotted in their figure 3: on 7th Jan 2001 $\tau=-94\pm4\;{\rm s}$ at UT12:14, and $\tau=+107\pm4\;{\rm s}$ at UT15:54; on 12th Jan 2001 $\tau=-76\pm13\;{\rm s}$ at UT11:58, and $\tau=+121\pm21\;{\rm s}$ at UT16:07. Here the sign convention is that a positive value of $\tau$ means that features in the light-curve occur earlier in the VLA data than in the WSRT data. As for PKS1257$-$326 we attempt to fit both types of data simultaneously and we do so with both one- and two-dimensional models. In fitting our models to the data, each of the four delay measurements was given equal weight with each of the 39 time-scale measurements. Our best-fit two-dimensional model proved to be unphysical, with a superluminal major-axis velocity component. An unphysical solution is of no interest so we proceeded by fixing the major-axis velocity at the largest physically plausible value, which we chose to be $400\;{\rm km\,s^{-1}}$ --- comparable to the escape speed from the Galaxy. The resulting best-fit model parameters are given in table 1, and the corresponding annual cycle in scintillation time-scale is shown in figure 2 for both best-fit models and the data. 

As was the case with PKS1257$-$326 we see that the best-fit  one- and two-dimensional models are very similar in respect of the values of the parameters which they have in common, and that they differ only slightly in respect of the predicted scintillation time-scales. The two-dimensional model matches the data a little better, as measured by the $\chi^2$ of the fits, but the reduced $\chi^2$ values
are $\chi_r^2 = 1.315$ for the one-dimensional model and $\chi_r^2=1.324$ for the two-dimensional model --- indicating that the difference in $\chi^2$ is not significant and the extra two parameters needed for the two-dimensional model are not justified by these data.

\begin{figure}
\vskip -1.5cm\hskip -1.5cm\includegraphics[width=11cm]{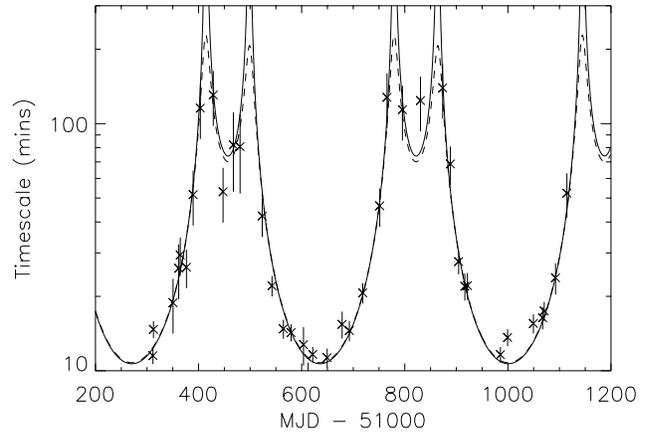}
\vskip -6.5cm
\caption{The annual cycle in scintillation time-scale for J1819$+$3845 for best-fit one- and two-dimensional models (solid and dashed lines, respectively), along with the data.}
\end{figure}

\subsection{Solution geometries}
Care is needed to avoid confusion in interpreting the models presented here because the major axis of the scintillation pattern defines our coordinate axes ($\parallel,\perp$) only up to an ambiguity of $180^\circ$, leaving a sign ambiguity in $v_\perp$. To resolve this ambiguity we present a vector diagram of  our one-dimensional best-fit models in figure 3.

Comparing our best-fit one-dimensional models with the two-dimensional models of Dennett-Thorpe \&\ de~Bruyn (2003) and Bignall et al (2003, 2006), we find that there is good agreement in respect of the orientation of the major axis and the value of $v_\perp$. This is all that can be expected since the results of \S2.1 and \S2.2 demonstrate that $R$ and $v_\parallel$ cannot be meaningfully constrained by the available data.

The velocity $\vec{v}$ which we have used to parameterise our models describes all contributions to the effective scintillation velocity, $\vec{V}$, other than that due to the Earth's orbit, $\vec{u}_\oplus$, which is included explicitly:  $\vec{V}=\vec{u}_\oplus + \vec{v}$. Quasars are so distant that we can neglect the contribution of the source to the effective scintillation velocity and $\vec{v}$ therefore represents the velocity of the Solar System barycentre relative to the velocity of the scattering screen. Following Dennett-Thorpe \&\ de~Bruyn (2003) we take the velocity of the Solar System barycentre relative to the Local Standard of Rest (LSR) to be $19.7\;{\rm km\,s^{-1}}$ toward ${\rm 18^h07^m50^s, +30^\circ00^\prime52^{\prime\prime}}$ (J2000). The contribution to $v_\perp$ arising from this motion is small ($2.0\;{\rm km\,s^{-1}}$) in the case of J1818$+$3845, but large ($17.2\;{\rm km\,s^{-1}}$) for PKS1257$-$326. Thus the similarity in our best-fit $v_\perp$ values for the two sources is coincidental.

\begin{figure}
\vskip -1.5cm\hskip -1.5cm\includegraphics[width=11cm]{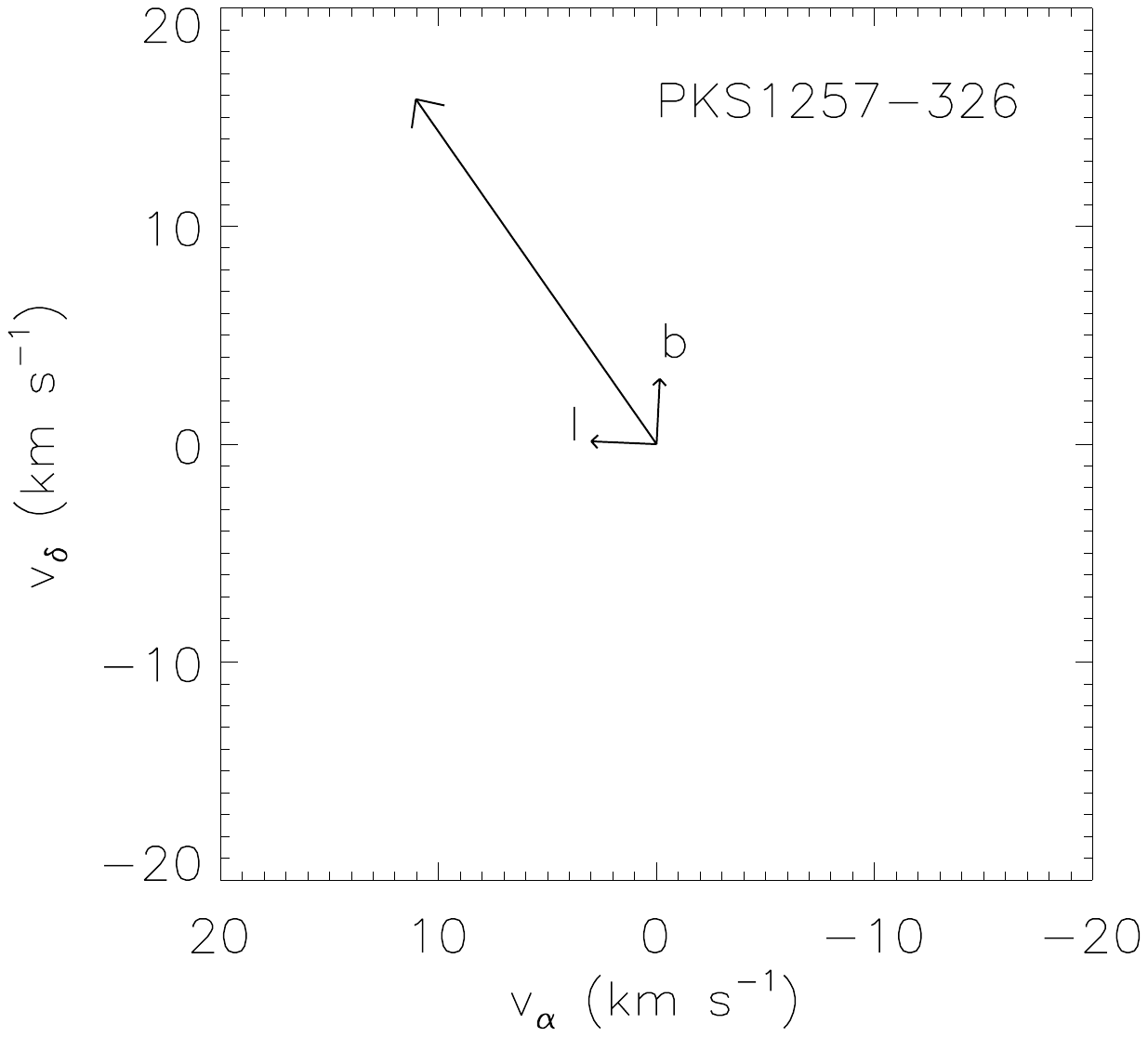}
\vskip -8.0cm\hskip -1.5cm\includegraphics[width=11cm]{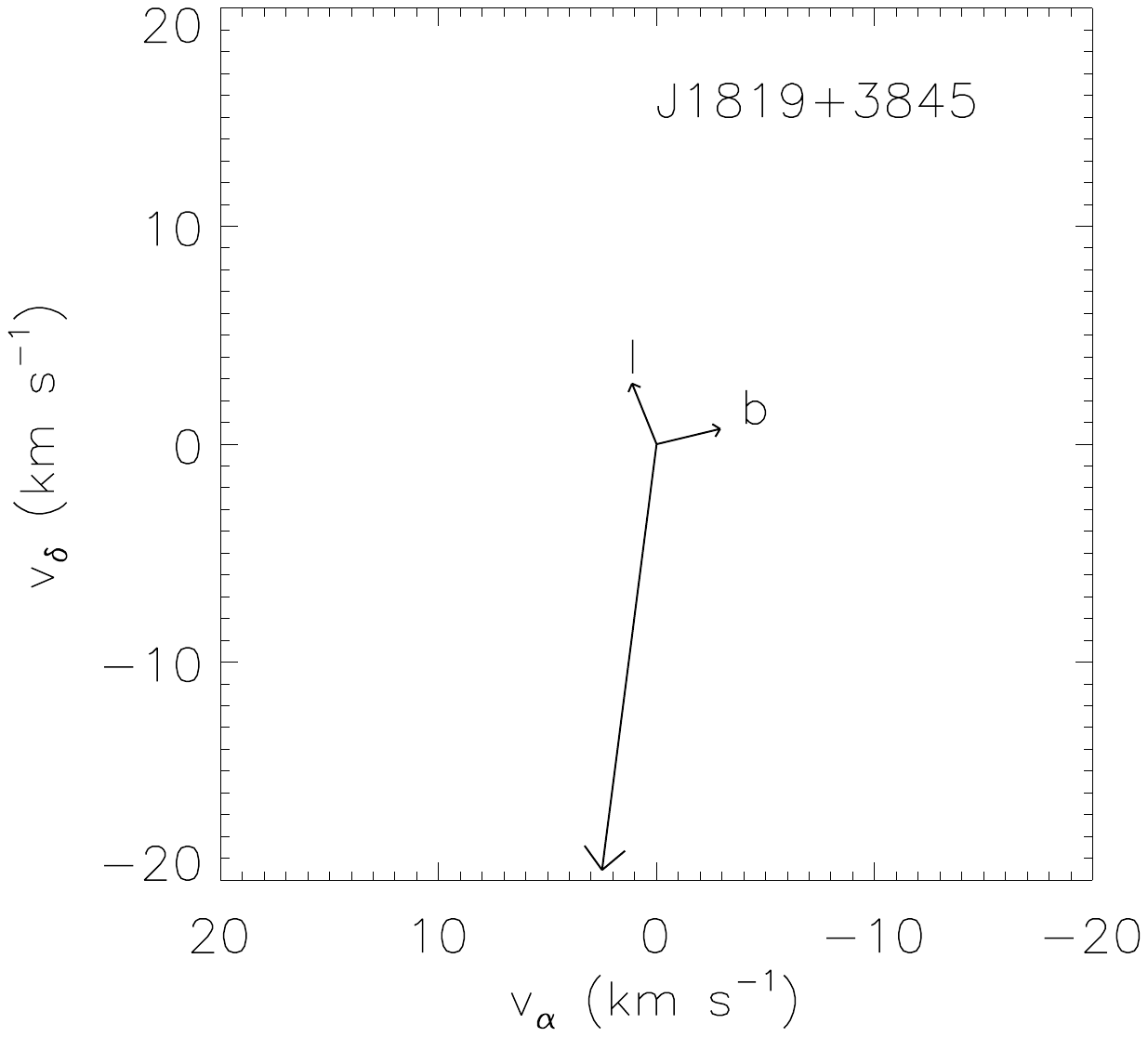}
\vskip -6.5cm
\caption{Magnitude and orientation of $v_\perp$ for PKS1257$-$326 (top panel) and J1819$+$3845 (bottom panel) as determined by the best-fit totally anisotropic scattering model in each case. Also plotted on each figure are vectors corresponding to the local  directions of increasing Galactic longitude ($l$) and latitude ($b$).}
\end{figure}

\section{Discussion}
The models employed in \S2 are descriptions of the geometry and kinematics of the scintillation pattern. We have just noted that the source is not expected to contribute significantly to the pattern velocity, but it could be important in determining the length-scales $a_\perp,a_\parallel$ because the observed pattern is a convolution of the point-source response of the screen with the (demagnified, inverted) source image. We must therefore consider whether the scintillation patterns for PKS1257$-$326 and J1819$+$3845 are quasi-one-dimensional because of the source structure or because of the phase structure in the scattering screen. Although extragalactic radio sources commonly exhibit elongated structure (``jets''), the axis ratios of their images are generally small in comparison with the pattern anisotropies which we are considering here. For example, in the well-studied case of the source 3C120 the jet length:width ratio is less than 10 almost everywhere in a logarithmically broad range of angular scales (Walker, Benson \&\ Unwin 1987), whereas the smallest pattern axis ratio given in our table 1 exceeds 100. By contrast the phase screen anisotropy is already known to be large in the case of the rapid scintillators PKS0405$-$385 and PKS1257$-$326 (Rickett, Kedziora-Chudczer \&\ Jauncey 2002; Bignall et al 2006), albeit with a modest lower limit ($R>4$).  Consequently it is natural to interpret our result as extreme anisotropy in the phase screen rather than in the source. In fact this interpretation is the only possible one because if $a_\parallel\rightarrow\infty$ as a result of a very elongated  source, then the amplitude of the resulting scintillations should be small, because of the smoothing effect of the convolution, whereas it is observed to be large. Thus the extreme pattern anisotropy which we infer requires extremely anisotropic scattering.

Why go to the bother of testing a one-dimensional scattering model when we know that the real scattering screens must have a finite anisotropy? The point here is one of scale: we do not know what the anisotropy is even to order of magnitude. It does however appear to be large: not only is the one-dimensional model entirely adequate to describe the data, but the best of all two dimensional models are those with anisotropy $R > 10^2$. Very large values of the anisotropy are consistent with the high level of spectral purity in the radio light-curves of the fast scintillators (Rickett, Kedziora-Chudczer \&\ Jauncey 2002), but that diagnostic test is insensitive to the precise value of $R$ once it becomes large. This raises the question: how can the anisotropy be measured? Figure 2 demonstrates that when the scintillation time-scale is very long it is sensitive to finite anisotropy, but unfortunately it is also difficult to measure accurately in this case. Indeed Dennett-Thorpe \&\ de~Bruyn (2003) show three light-curves (recorded on 27/08/1999, 28/11/1999 and 27/08/2000) for which the variations are extremely slow and no measurement of scintillation time-scale is reported in their table 3. We now describe an alternative approach to measuring $R$ which can be employed for J1819$+$3845.

The scintillation time-scale is formally infinite in the one-dimensional scattering model at times when $V_\perp=0$. (According to our best-fit totally anisotropic scattering model for J1819$+$3845 this condition is satisfied each year sometime around 23rd August and again around 15th November.) These points correspond to changes in the sign of $V_\perp$ and they are therefore turning-points, $x_{\perp, min}$, $x_{\perp, max}$, in the minor-axis coordinate (which we shall denote $x_\perp$). This is illustrated in figure 4 where we see that each value of $x_{\perp}$ in the range $x_{\perp, min} < x_{\perp} < x_{\perp, max}$ occurs at three distinct times each year --- once between the two times when $V_\perp$ changes sign, once before this interval and once after.  In the totally anisotropic model the flux depends only on $x_\perp$ so we expect a one-to-one correspondence between fluxes measured before and after the turning points, where those measurements are made at the same value of $x_{\perp}$.

In practice the anisotropy, $R$, must be finite albeit large. A consequence of this is that a pair of points with the same value of $x_{\perp}$ but having major-axis coordinate separations $|\delta x_\parallel| \ga a_\parallel \simeq 30\,R\;{\rm Mm}$ will exhibit significantly different fluxes. The correlation between fluxes taken pair-wise in this way increases as  $|\delta x_\parallel|$ decreases and is very strong for $|\delta x_\parallel| \ll a_\parallel$.  In this way we can measure $a_\parallel$ if we know the trajectory $x_\parallel$ as a function of time, which in turn is completely determined by the value of $v_\parallel$.   We do not know $v_\parallel$ independently but its value can be determined simultaneously with that of $a_\parallel$ because $V_\parallel$ at the two turning points differs by about $40\;{\rm km\,s^{-1}}$ as a result of the different contributions from the Earth's velocity vector.

In figure 4 we can find, on each trajectory, separations of up to $|\delta x_\parallel | \sim 10^5\;{\rm Mm}$ between points with the same perpendicular coordinate, indicating that for these trajectories we can estimate the major-to-minor axis ratio if its value is $R\la3\times10^3$. Figure 4 shows three example trajectories, all of which have $|v_\parallel | \le 15\; {\rm km\,s^{-1}}$; if, on the other hand, the magnitude of the parallel velocity component is as great as $400\;{\rm km\,s^{-1}}$ then this approach is sensitive to still larger values of the anisotropy: $R\la 10^5$. These estimates have all been carried through in the ``frozen screen'' approximation, for which it is meaningful to assign a unique steady velocity $\vec{v}$ to the scintillation pattern. The fact that we are able to find good fits to the existing time-scale and time-delay data for PKS1257$-$326 and J1819$+$3845 demonstrates that the model is indeed a valid description of the perpendicular motion, but it might fail on the longer time-scales needed to explore the scintillation pattern along the major axis direction. Departures from the simple kinematic model we have used would provide constraints on any physical model of the scattering media.

\begin{figure}
\vskip -1.5cm\hskip -1.15cm\includegraphics[width=11cm]{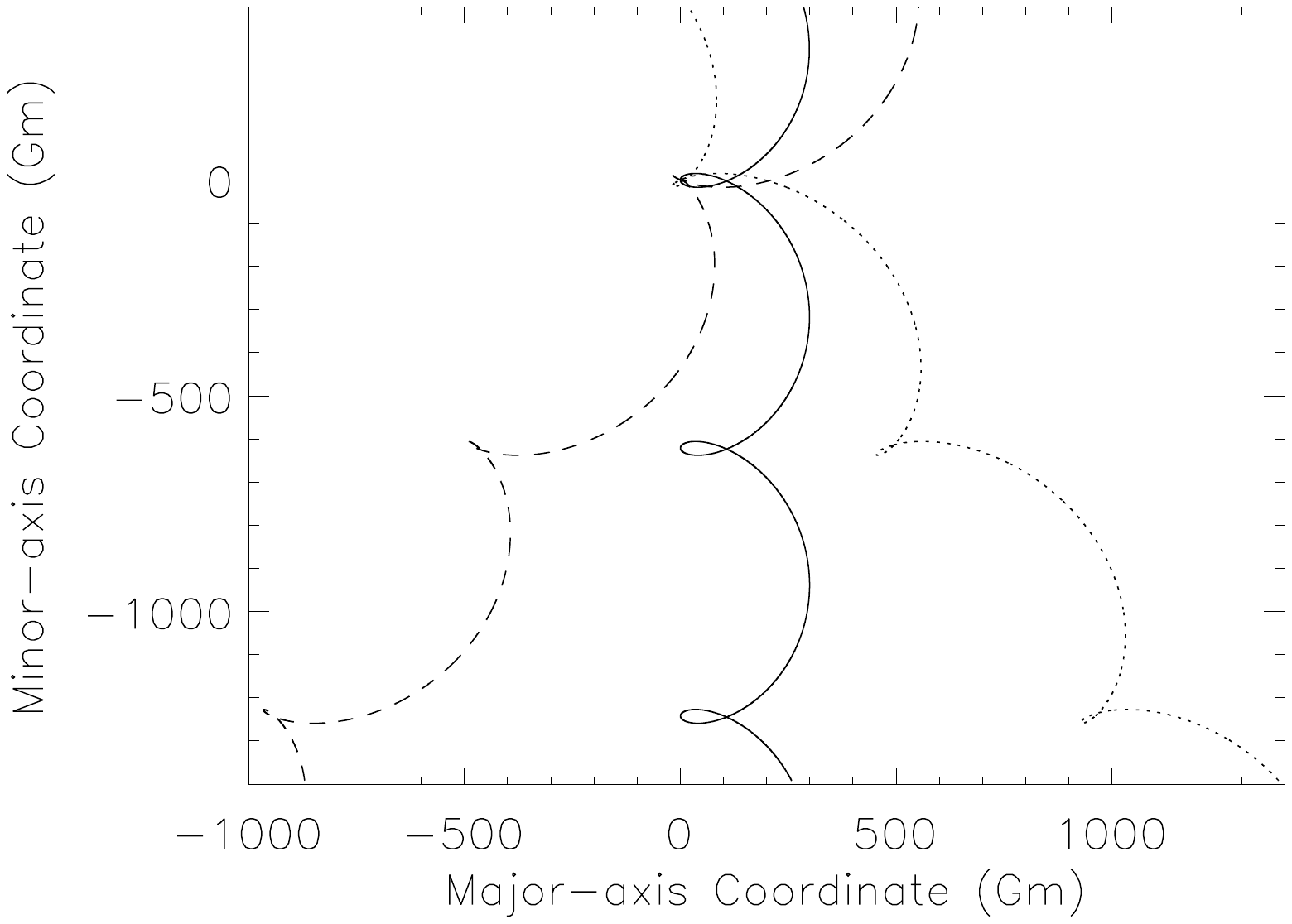}
\vskip -10.5cm\hskip -1.5cm\includegraphics[width=11cm]{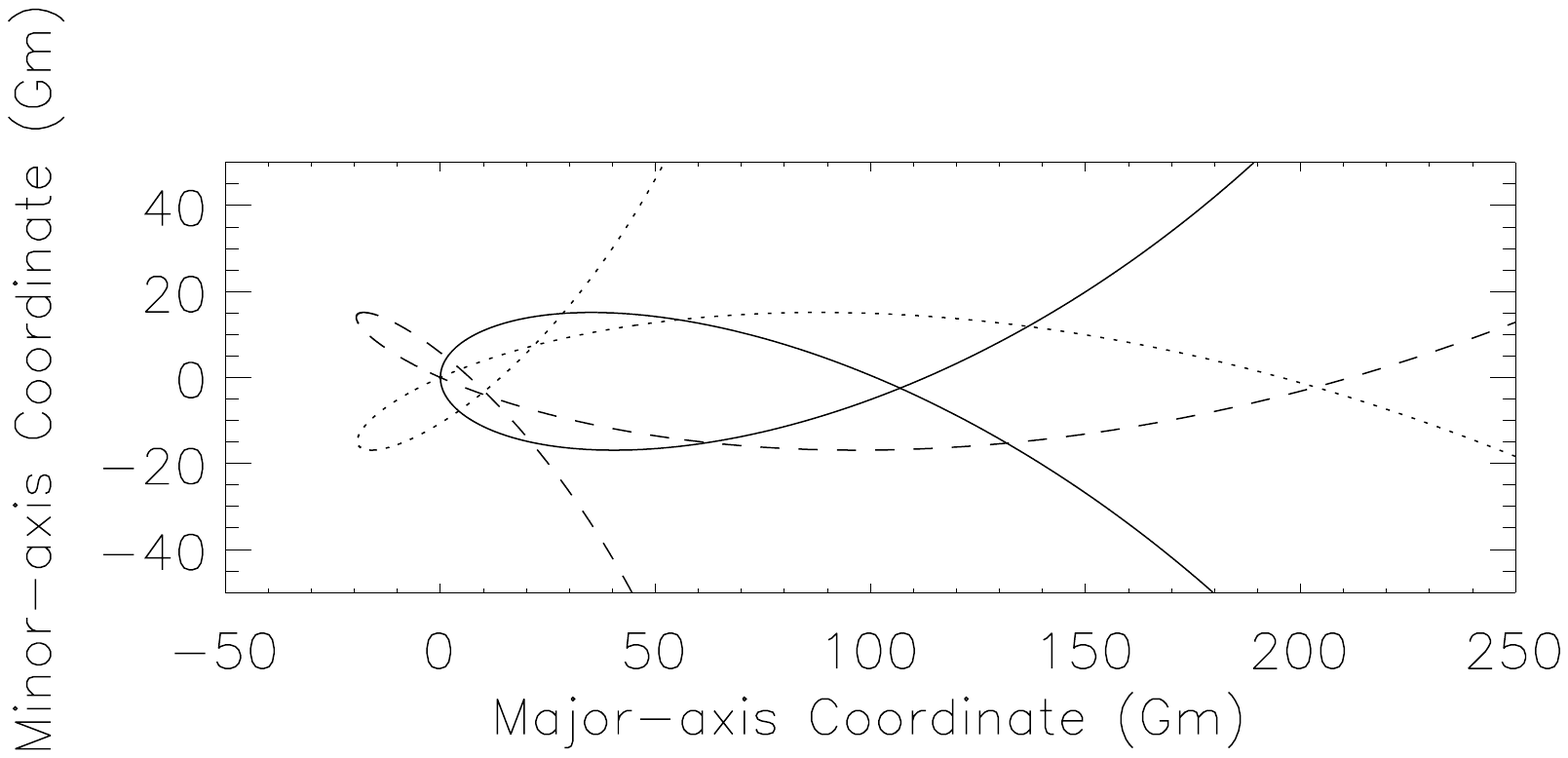}
\vskip -6.5cm
\caption{Three possible trajectories of an observer relative to the J1819$+$3845 scintillation pattern over a roughly three-year interval (top panel): the solid line corresponds to $v_\parallel=0$; dotted and dashed lines correspond to $v_\parallel=\pm15\;{\rm km\,s^{-1}}$. The origin of coordinates is arbitrary for each curve.  Turning points in the minor-axis coordinate correspond to $V_\perp=0$ and this condition is satisfied on the same day regardless of $v_\parallel$. These turning points occur in late August and mid November each year. Each value of the minor-axis coordinate between the turning points occurs on three separate occasions for all possible trajectories. The lower panel shows a close-up of the region around the origin of coordinates in the top panel.}
\end{figure}

\section{Conclusions}
At present there is no evidence for finite anisotropy in the scattering screens responsible for the intra-hour variations of PKS1257$-$326  and J1819$+$3845 and only a one-dimensional model is appropriate for each source.  In the case of J1819$+$3845 a strictly one-dimensional model predicts that the particular  scintillations seen between late August and mid November each year should also be seen before and after this period as the Earth's velocity across the flux pattern changes sign at these times and the observer moves back and forth across the same part of the pattern.  If repeating flux patterns can be identified in the data they will permit precise determination of the major-axis orientation and the minor-axis velocity component. To the extent that decorrelation of the pair-wise fluxes can be quantified we can measure the major-to-minor axis ratio, $R$, and the major-axis velocity component of the screen. Decorrelation is expected to be observable only if $R\la 10^5$.

\section{Acknowledgments}
MAW thanks J.P.~Macquart for prompting this study through his scepticism that a one-dimensional model could fit the data for J1819$+$3845.

\end{document}